\documentclass[reprint,superscriptaddress,amsmath,amssymb]{revtex4-2}

\usepackage{graphicx}
\usepackage{dcolumn}
\usepackage{bm}
\usepackage{braket}
\usepackage{appendix}
\usepackage{subfigure}
\usepackage{hyperref}
\usepackage{color}
\usepackage{xcolor}
\usepackage{amsmath}
\usepackage{nccmath}
\usepackage[english]{babel}
\usepackage{svg}
\usepackage{multirow}
\usepackage{tabularx}
\usepackage{array}
\usepackage{blindtext}
\usepackage{titlesec}
\usepackage{soul}
\usepackage{ulem}
\newcolumntype{P}[1]{>{\centering\arraybackslash}p{#1}}

\usepackage{lineno}

\begin{document}

	\title{Single nuclear spin detection and control in a van der Waals material}
		
	\author{Xingyu Gao}
	\thanks{These authors contributed equally to this work.}
	\affiliation{Department of Physics and Astronomy, Purdue University, West Lafayette, Indiana 47907, USA}

  \author{Sumukh Vaidya}
  \thanks{These authors contributed equally to this work.}
\affiliation{Department of Physics and Astronomy, Purdue University, West Lafayette, Indiana 47907, USA}

\author{Kejun Li}
\affiliation{Department of Physics, University of California, Santa Cruz, CA, 95064, USA}
\affiliation{Department of Materials Science and Engineering, University of Wisconsin-Madison, 53706, USA}

\author{Zhun Ge}
\affiliation{Department of Physics and Astronomy, Purdue University, West Lafayette, Indiana 47907, USA}

\author{Saakshi Dikshit}%
\affiliation{Elmore Family School of Electrical and Computer Engineering, Purdue University, West Lafayette, Indiana 47907, USA}

\author{Shimin Zhang}
\affiliation{Department of Materials Science and Engineering, University of Wisconsin-Madison, 53706, USA}

\author{Peng Ju}
\affiliation{Department of Physics and Astronomy, Purdue University, West Lafayette, Indiana 47907, USA}

\author{Kunhong Shen}
\affiliation{Department of Physics and Astronomy, Purdue University, West Lafayette, Indiana 47907, USA}

 \author{Yuanbin Jin}

\affiliation{Department of Physics and Astronomy, Purdue University, West Lafayette, Indiana 47907, USA}

\author{Yuan Ping}
\affiliation{Department of Materials Science and Engineering, University of Wisconsin-Madison, 53706, USA}
\affiliation{Department of Physics, University of Wisconsin-Madison, 53706, USA}
\affiliation{Department of Chemistry, University of Wisconsin-Madison, 53706, USA}

\author{Tongcang Li}%
\email{tcli@purdue.edu}
\affiliation{Department of Physics and Astronomy, Purdue University, West Lafayette, Indiana 47907, USA}
\affiliation{Elmore Family School of Electrical and Computer Engineering, Purdue University, West Lafayette, Indiana 47907, USA}
\affiliation{Purdue Quantum Science and Engineering Institute, Purdue University, West Lafayette, Indiana 47907, USA}
\affiliation{Birck Nanotechnology Center, Purdue University, West Lafayette, Indiana 47907, USA}
\date{\today}

\begin{abstract}
{\normalsize 
Optically active spin defects in solids \cite{atature2018material,wolfowicz2021quantum} are leading candidates for quantum sensing \cite{degen2017quantum,du2024single} and quantum networking \cite{pompili2021realization,knaut2024entanglement}. Recently, single spin defects were discovered in hexagonal boron nitride (hBN) \cite{chejanovsky2021single,mendelson2021identifying,stern2022room,guo2023coherent,stern2023quantum}, a layered van der Waals (vdW) material. Due to its two-dimensional structure, hBN allows spin defects to be positioned closer to target samples than in three-dimensional crystals, making it ideal for atomic-scale quantum sensing \cite{shen2024proximity}, including nuclear magnetic resonance (NMR) of single molecules. However, the chemical structures of these defects \cite{chejanovsky2021single,mendelson2021identifying,stern2022room,guo2023coherent,stern2023quantum} remain unknown, and detecting a single nuclear spin with an hBN spin defect has been elusive. In this study, we created single spin defects in hBN using $^{13}$C ion implantation and identified three distinct defect types based on hyperfine interactions. We observed both  $S=1/2$ and $S=1$ spin states within a single hBN spin defect. We demonstrated atomic-scale NMR and coherent control of individual nuclear spins in a vdW material, with a $\pi$-gate fidelity up to 99.75\% at room temperature. By comparing experimental results with density-functional theory calculations, we propose chemical structures for these spin defects. Our work advances the understanding of single spin defects in hBN and provides a pathway to enhance quantum sensing using hBN spin defects with nuclear spins as quantum memories.
}
\end{abstract}

\maketitle


\vspace{0.1in}



Solid-state spin defects have become a leading platform for a wide range of quantum technologies, including  multinode quantum networking \cite{pompili2021realization,knaut2024entanglement} and quantum-enhanced sensing \cite{degen2017quantum,du2024single}. These advances are largely driven by the spin-photon quantum interfaces, which utilize optically addressable coherent spins. Despite the success of various spin-photon systems, each material platform has intrinsic limitations, creating trade-offs depending on the specific application \cite{atature2018material,wolfowicz2021quantum}. Moreover, spin defects that operate at room temperature are rare. 
Recently, optically active spin defects in hexagonal boron nitride (hBN), a van der Waals (vdW) material, has garnered vast attention \cite{gottscholl2020initialization,chejanovsky2021single,mendelson2021identifying,stern2022room,guo2023coherent,stern2023quantum}. The layered structure of hBN facilitates integration with nanophotonic devices \cite{caldwell2019photonics} and provides an ideal platform for quantum sensing at the atomic scale \cite{shen2024proximity}. hBN spin defects have been used for sensing magnetic fields \cite{gottscholl2021spin,healey2022quantum,huang2022wide}, temperature \cite{healey2022quantum,gottscholl2021spin}, strain \cite{lyu2022strain}, and  beyond \cite{vaidya2023quantum}. 
 However, electron spins in hBN suffer from short spin coherence times \cite{gottscholl2021room,gong2023coherent,rizzato2023extending}. 

Nuclear spins typically exhibit long coherence times thanks to their weak coupling with the local environment, making them ideal candidates for quantum registers \cite{dutt2007quantum}.  By employing nuclear spins as ancillary qubits, we can overcome the limitation of electron spin coherence times and enhance the sensitivity of a spin-based quantum sensor \cite{zaiser2016enhancing,aslam2017nanoscale}. This approach requires the ability to initialize, control, and read out individual nuclear spins \cite{neumann2010single,waldherr2014quantum}. In the context of hBN spin defects, previous experiments used negatively charged boron vacancy ($V_B^-$) defects to polarize and read out nitrogen nuclear spin ensembles \cite{gao2022nuclear,gong2024isotope,clua2023isotopic,ru2024robust}. However,  these experiments were limited to ensemble-level operations due to the low quantum efficiency of $V_B^-$ defects. Moreover, the relatively short relaxation time of $V_B^-$ electron spins ($T_1$ $<$ 20 $\mu$s at room temperature)  restricted the maximum operational time of nuclear spins. 
Recently discovered single spin defects in hBN \cite{chejanovsky2021single,mendelson2021identifying,stern2022room,guo2023coherent,stern2023quantum} enabled the readout of individual electron spins in hBN. However, the chemical structures of these single spin defects remain unidentified, and the control of individual nuclear spins in hBN or other vdW materials is still elusive.

\begin{figure*}[tph]
	\centering
	\includegraphics[width=0.95\textwidth]{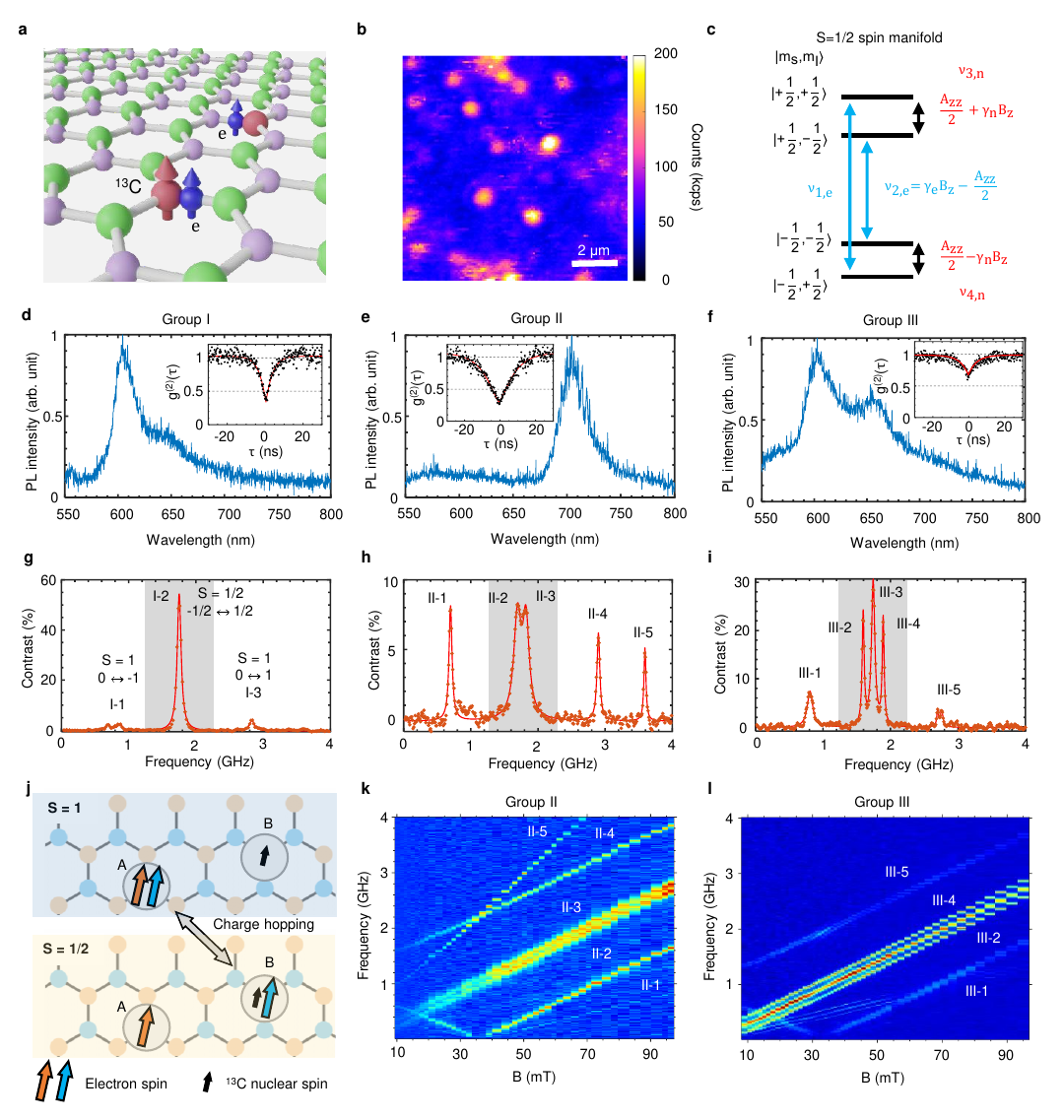}
	\caption{\textbf{Observation of three types of single spin defects in hBN.}  (a) An illustration of a carbon-related spin defect complex, consisting of an electron spin strongly coupled to a $^{13}$C nuclear spin and a nearby, weakly coupled electron spin without strong hyperfine interaction. (b)  A PL confocal map showing isolated bright emitters in hBN. Scale bar: 2 $\mu$m. (c) An energy level diagram of an electron spin $S=1/2$, coupled to a $^{13}$C nuclear spin ($I=1/2$). $A_{zz}$ is the hyperfine interaction strength. $\nu_{1,e}$ and $\nu_{2,e}$ are the two electron spin transitions. (d)-(f) Optical spectra of Defect 1-3, belonging to Group I-III, respectively. (g)-(i) ODMR spectra of Defects 1–3 under an out-of-plane external magnetic field of 62.5 mT.  The number of peaks in the central region (shaded area) differs among Group I–III defects.  Within this shaded area, the peaks correspond to the $\ket{-1/2}$ $\leftrightarrow$ $\ket{+1/2}$ transitions, with hyperfine structures observed in Groups II and III. Outside the shaded region, the transitions correspond to the $\ket{0}$ $\leftrightarrow$ $\ket{\pm1}$ transitions. Additionally, the defect in (h) exhibits a $\ket{-1}$ $\leftrightarrow$ $\ket{+1}$  double-quantum transition (II-5). (j) An illustration of the spin pair model for explaining the coexistence of S = 1 and S = 1/2 transitions.  (k)-(l) Magnetic field dependent ODMR spectra of  Defect 2 in Group II (k), and  Defect 3 in Group III (l).} \label{fig:1}
\end{figure*}

In this article, we report the first realization of single nuclear spin detection and control using a carbon-related defect in hBN. Our study reveals three major types of single spin defects in $^{13}$C implanted hBN, with remarkably high optically detected magnetic resonance (ODMR) contrasts of up to 200$\%$. Two of these defect types are strongly coupled to nearby $^{13}$C nuclear spins. We observe the coexistence of $S=1$ and $S=1/2$ states within a single spin defect, though only the $S=1/2$ states exhibit strong coupling to $^{13}$C nuclear spins, with coupling strengths reaching up to 300 MHz. The strong hyperfine coupling results in well resolved hyperfine structures, enabling the initialization and readout of single nuclear spins assisted by a defect electron spin. We realize electron-nuclear spin two-qubit gate and  coherent control of a  $^{13}$C nuclear spin, with a nuclear spin $\pi$-gate fidelity up tp 99.75\% at room temperature. We also perform Ramsey and Hahn Echo measurements of the nuclear spin. The dephasing and coherence times of a typical $^{13}$C nuclear spin in hBN at room temperature are measured to be $T_2^*$ = 16.6 $\mu$s and $T_2$ = 162 $\mu$s, respectively. Our density-functional theory calculations suggest that C$_B^+$C$_N^0$ donor-acceptor pairs (DAP) and C$_B$O$_N$ are likely key components for the two types of spin defects with large hyperfine interactions.

\section*{Single spin defects in $^{13}$C implanted hBN}

We create carbon-related spin defects in hBN \cite{mendelson2021identifying,stern2022room,stern2023quantum}  (Fig. \ref{fig:1}) by $^{13}$CO$_2$ (99$\%$ $^{13}$C) ion implantation and thermal annealing (Supplemental Section II). A confocal photoluminescence (PL) map reveals isolated emitters, as depicted in Fig. \ref{fig:1}(b). Their optical spectra spread from 570 nm to 700 nm, depending on different defects (Figure \ref{fig:1}(d)-(f)). Photon correlation measurements suggest that some defects are single photon emitters ($g^{(2)}(0)<0.5$). To verify their spin properties, we perform ODMR measurements by toggling the microwave on and off while collecting the emitted photons. The ODMR contrast $C$ is determined by the ratio of photon count rates when the microwave is on ($N_{\rm on}$) or off ($N_{\rm off}$): $C=({N_{\rm on}-N_{\rm off}})/{N_{\rm off}}\times100\%$. In contrast to previous reports of $S=1/2$ spin defects with only a single resonant peak in ODMR \cite{chejanovsky2021single,mendelson2021identifying,stern2022room,guo2023coherent}, we observe three distinct ODMR spectra with multiple resonances (Fig. \ref{fig:1}(g)-(i)). The positive contrast indicates that the defects are initialized into a darker state, while we also observe defects with negative contrasts but are rarer (Supplemental Figure S20). These ODMR spectra feature a center branch (shaded areas in Fig. \ref{fig:1}(g)-(i)) comprising multiple resonances that are assigned to hyperfine structures, along with additional side resonances located approximately 1 GHz away from the center branch. The high ODMR contrast and brightness yield a typical DC magnetic field sensitivity of 5~$\mu$T/$\sqrt{\rm Hz}$ (Methods).

Based on ODMR spectra as shown in Fig. \ref{fig:1}(g)-(i), we categorize the spin defects into three major groups, Group I-III, according to the number of peaks in the center branch. Individual peaks are labeled according to their respective defect group. The hyperfine splitting of Group II and III defects are determined to be 130 MHz and 300 MHz, respectively. Notably, the 300 MHz hyperfine splitting is determined by the separation between transitions III-2 and III-4. The additional center peak, III-3, exhibits a different hyperfine stucture under weak microwave driving, suggesting that it originates from a second electron spin.  This observation supports the recently proposed spin-pair model, in which a single defect complex hosts two electron spins, separated by several nanometers, that couple differently to the $^{13}$C nuclear spins (Fig. \ref{fig:1}(j)) \cite{robertson2024universal,patel2024room}.

To better understand the spin transitions, we measure the ODMR spectra as a function of out-of-plane magnetic fields (Figure \ref{fig:1}(k) and \ref{fig:1}(i)). The results reveal an absence of zero-field splitting (ZFS) in the center branch, corresponding to the $\ket{m_s=+1/2}$ $\leftrightarrow$ $\ket{m_s=-1/2}$ transition within the $S=1/2$ spin manifold (see Supplemental Section IV for experimental verification). In contrast, the side resonances (II-1, II-4, III-1 and III-5) exhibit non-zero ZFS and disperse with a $g$-factor of 2, confirming that they correspond to the $\ket{m_s=0}$ $\leftrightarrow$ $\ket{m_s=\pm1}$ transitions within the $S = 1$ spin manifold. The Hamiltonian of each spin manifold is given by:
\begin{equation}
	H = DS_z^2+E(S_x^2-S_y^2)+\gamma_e \mathbf{B}\cdot \mathbf{S}+\sum_{i}\mathbf{S}\cdot A_i \cdot \mathbf{I_i}+\gamma_n \mathbf{B}\cdot\mathbf{I_i}, \label{Eq:H_spin}
\end{equation} 
where $\gamma_e$ and $\gamma_n$ are the electron spin and the nuclear spin gyromagnetic ratios, and $\mathbf{B}$ is the external magnetic field. $\mathbf{S}$ denotes the electronic spin operator with $S = 1$ or $1/2$. $D$ and $E$ together are the ZFS parameters, which are non-zero for the $S=1$ states  but vanish for the $S = 1/2$ states.  $\mathbf{I_i}$ denotes the nuclear spin operator for the $i$-th nuclear spin coupled to the defect electron. Specifically, $I = 1/2$ for $^{13}$C nuclear spins, $I = 1$ for $^{14}$N nuclear spins, and $I = 3/2$ for $^{11}$B nuclear spins. The hyperfine interaction between the electron spin and the i-th nuclear spin is described by the interaction tensor $A_{i}$.
 
The magnetic field dependent ODMR allows us to extract the ZFS parameter $D$ (typically $\sim$ 1 GHz) and $E$ (varies from 100 MHz to 400 MHz) along an out-of-plane quantization axis for the S = 1 transitions (Supplementary Figure S3).
This is different from the recent observation of a spin-triplet defect with an in-plane quantization axis \cite{stern2023quantum}, making our system more favorable for magnetic field alignment. Additionally, an extra peak II-5 of Defect 2  locates at the frequency $\nu_5$ = $\nu_1$ $\pm$ $\nu_4$ and has an LAC with II-4 at 37 mT (Supplemental Figure S3), suggesting a double-quantum transition. While previously proposed spin-pair models based on DAP have focused solely on S = 1/2 transitions in metastable state (MS) \cite{robertson2024universal, patel2024room}, the coexistence of both S = 1/2 and S = 1 transitions can be well explained by an extended spin-pair model incorporating additional energy levels and transitions (see Methods). The DAP model also accounts for the large variation in PL spectra observed across different defects, despite their similar ODMR signatures \cite{tan2022donor, pelliciari2024elementary}.

\begin{figure*}[tph]
	\centering
	\includegraphics[width=0.99\textwidth]{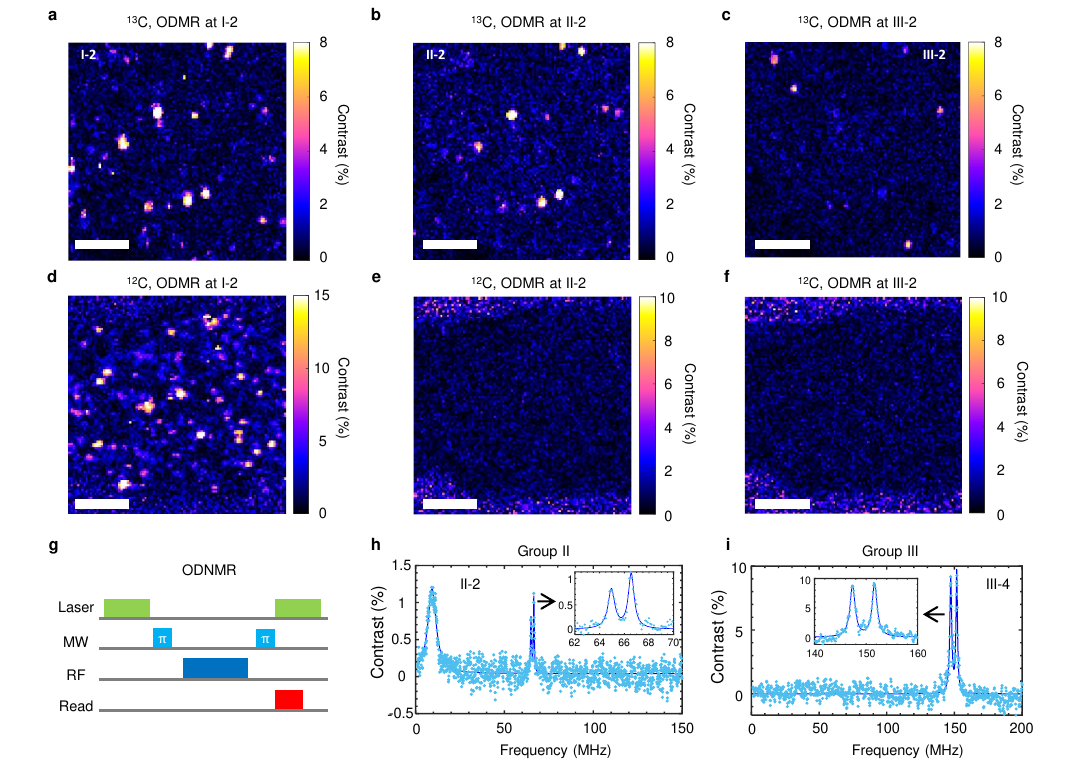}
	\caption{ \textbf{Optical detection of $^{13}$C nuclear spins in hBN.}  (a)-(c) ODMR contrast map of $^{13}$C implanted hBN by driving a microwave at (a) 2.01 GHz, resonance I-2 of Group I defects; (b) 1.95 GHz, resonance II-2 of Group II defects; and (c) 1.86 GHz, resonance III-2 of Group III defects. (d)-(f) ODMR contrast map of $^{12}$C implanted hBN by driving the microwave at (d) I-2, (e) II-2, and (f) III-2. The contrast fluctuation outside the hBN is caused by the low photon counts collected from the background. Scale bars are 5 $\mu$m. A 71.5 mT magnetic field is applied out-of-plane (perpendicular to the hBN nanosheet). (g) An illustration of the ODNMR sequence. (h) An ODNMR spectrum of Defect 2 by driving the microwave at II-2. (i) An ODNMR spectrum of Defect 3 by driving the microwave at III-4.  } \label{fig:2}
\end{figure*}

\section*{Detection of single $^{13}$C nuclear spins in hBN}

The well-resolved hyperfine structures enable us to distinguish different defect groups through two-dimensional (2D) mapping of the ODMR contrast at specific microwave driving frequencies. Figure \ref{fig:2} (a)-(c) present the ODMR contrast distribution of a $^{13}$C-implanted hBN with microwaves resonant at I-2, II-2 and III-2 transitions (Fig. \ref{fig:1}(g)-(i)), respectively. All three maps reveal multiple spots with finite contrasts, enabling us to locate Group II and Group III defects. In contrast, no ODMR signal is observed in $^{12}$C-implanted hBN when microwaves are applied at the II-2 and III-2 transitions.

Besides ODMR, optically detected nuclear magnetic resonance (ODNMR) (Figure \ref{fig:2}(g))  of $^{13}$C nuclear spins can provide further insights into the electron-nuclear hyperfine coupling. Here we conduct these measurements at the center branch of Group II and III defects. For both Defect 2 and Defect 3, we observe two closely spaced resonances in ODNMR at approximately half the frequencies of the hyperfine splitting observed in ODMR (Figure \ref{fig:2}(h)-(i)). Such a feature verifies the $S=1/2$ configuration of the defect electron spin for these transitions. For a $S=1$ configuration, we expect the ODNMR resonance frequency to be the same as the hyperfine splitting in ODMR. The two-peak structure originates from the external magnetic field and/or a weaker hyperfine coupling to a second electronic spin. We also notice that different nuclear resonance frequencies are obtained when the microwave is applied at III-2 (III-4) and III-3 (Supplemental Figure S17).  These features further verify two $S=1/2$ electron spins are involved in Defect 3.

\begin{figure*}[tph]
	\centering
	\includegraphics[width=0.99\textwidth]{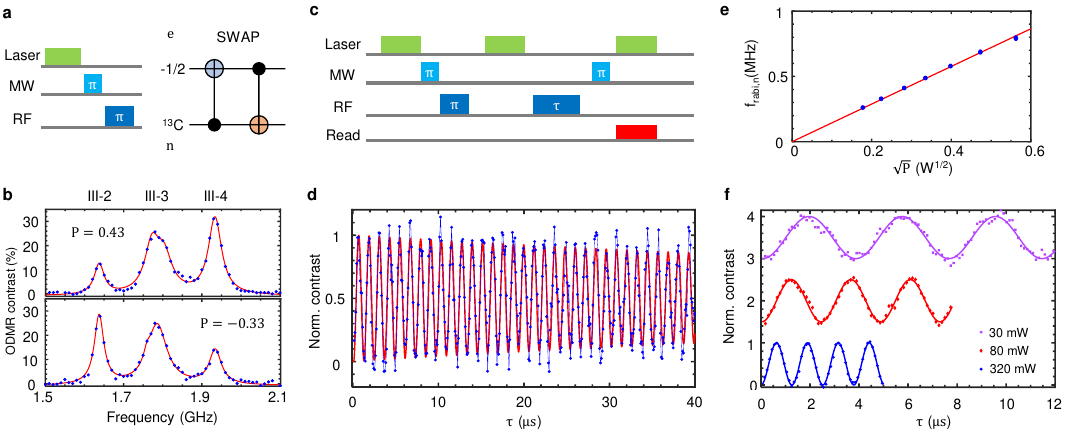}
	\caption{\textbf{Initialization and coherent control of a $^{13}$C nuclear spin.} (a) An illustration of the pulse sequence (left panel) for nuclear spin initialization. We use a SWAP gate (right panel) to transfer the electron spin polarization to the $^{13}$C nuclear spin. (b)  ODMR signal after nuclear spin initialization. One of the two peaks (III-2 and III-4) dominates, yielding a nuclear spin polarization of 0.43 (top panel) and -0.33 (bottom panel), depending on the direction of initialization. (c) Pulse sequence for nuclear spin coherent control.  (d) An example of nuclear spin Rabi oscillation, persisting for 40 $\mu$s without significant decay. The blue curve is the experimental data, and the red curve is fitting. (e) Nuclear spin Rabi frequency as a function of the square root of the RF power. (f) Nuclear spin Rabi oscillations taken at different RF powers: 30 mW (purple squares), 80 mW (Red diamonds), and 320 mW (Blue dots).} \label{fig:3}
\end{figure*}

\section*{Coherent control of a $^{13}$C nuclear spin in hBN}

\begin{figure*}[tph]
	\centering
	\includegraphics[width=0.65\textwidth]{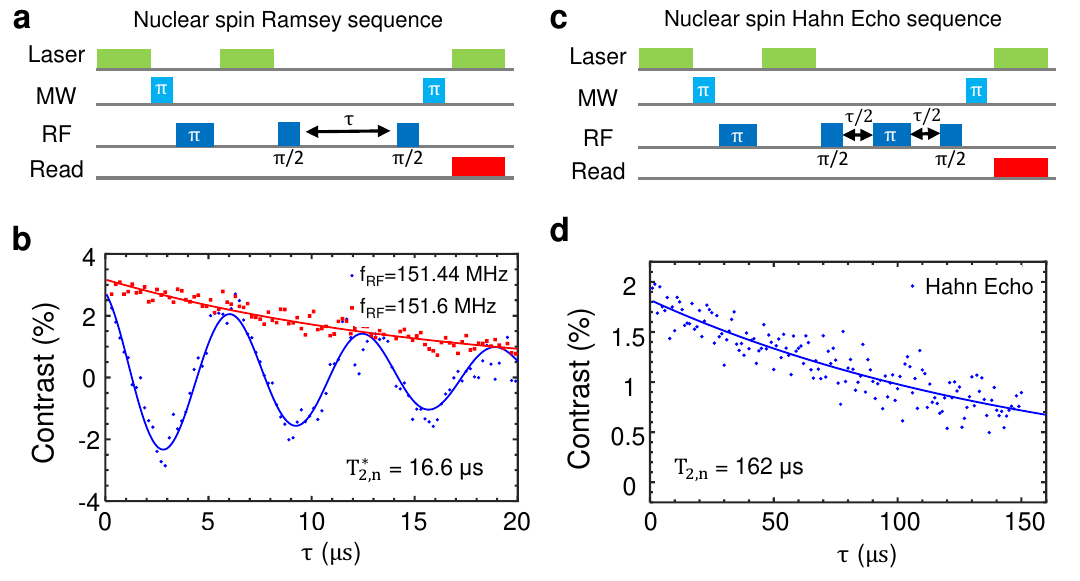}
	\caption{\textbf{Spin coherence of a $^{13}$C nuclear spin.} (a) The pulse sequence of nuclear spin Ramsey  interferometry. (b) The nuclear spin Ramsey fringe. The measurements are performed when the RF frequency is in resonance with the nuclear spin transition (red squares) or slightly detuned from the resonance (blue dots). The fitting of the oscillation showing an inhomogeneous dephasing time of $T_{2,n}^*$ = 16.6 $\mu$s.  (c) The pulse sequence of nuclear spin Hahn Echo. (d) The nuclear spin Hahn Echo measurement shows a slow decay with $T_{2,n}$ = 162 $\mu$s.} \label{fig:4}
\end{figure*}

The well-isolated electron spin transitions at III-2 and III-4 (Fig. \ref{fig:1}(i)) enable coherent manipulation of individual electron spin states associated with different nuclear spin states. This capability allows for the initialization, control, and readout of a nuclear spin via the electron spin, without the need for strong magnetic fields or level anticrossing (LAC) \cite{gao2022nuclear,gong2024isotope,clua2023isotopic,ru2024robust}. 

To polarize the $^{13}$C nuclear spin, we first use a laser pulse to initialize the electron state (assuming $\ket{m_s=-1/2}$).  Next, we
apply a SWAP gate (Figure \ref{fig:3}(a)) to swap the electron spin state and nuclear spin state, after which the nuclear spin is initialized to $\ket{\uparrow}$ (or $\ket{\downarrow}$). To estimate the fidelity of nuclear spin initialization, we measure the electron spin ODMR signal after the SWAP gate and calculate the polarization according to the imbalance between III-2 and III-4 (Figure \ref{fig:3}(b)). We determine the nuclear spin polarization to be approximately 43$\%$ (-33$\%$) using the equation $P=(\rho_4-\rho_2)/(\rho_4+\rho_2)$, where $\rho_j$ represents the amplitude of resonance III-j. A higher nuclear spin polarization of 60$\%$ can be achieved with another defect as shown in Supplemental Figure S18.

For nuclear spin coherent control, we use the protocol depicted in Figure \ref{fig:3}(c).  After polarizing the nuclear spin, another 10-$\mu$s laser pulse re-initializes the electron spin state. Subsequently, we park the frequency of the selective RF pulse while varying its pulse duration $\tau$. Finally, the nuclear spin state is read out via the spin defect using a selective microwave $\pi$ pulse and a 5-$\mu$s laser pulse. Figure \ref{fig:3}(d) shows the resulting nuclear spin Rabi oscillations. We determine a $\pi$-pulse time of 0.6 $\mu$s and a coherence time of 117 $\mu s$ by fitting the oscillation contrast $C(\tau)$ to $C(\tau)= a\cdot \sin(\pi\tau/T_{\pi}+b)\exp(-\tau/T_{Rabi})+d$, yielding a $\pi$-gate fidelity of 99.75\% (Methods). The extended operational time for nuclear spins is attributed to the long electron spin relaxation time ($T_{1,e}$ = 144 $\mu$s, Supplemental Figure S15) of this carbon-related spin defect, which is an order of magnitude longer than that of $V_B^-$ spin ensembles \cite{gao2022nuclear,gong2024isotope}. By repeating the measurements at different RF powers, we observe a clear power dependence of the oscillations (Figure \ref{fig:3}(f)), where the nuclear Rabi frequency is linearly proportional to the amplitude of the RF field (Figure \ref{fig:3}(e)).

We further characterize the nuclear spin coherence via Ramsey and Hahn Echo sequences, as illustrated in Figure \ref{fig:4} (a)-(b). In the Ramsey interferometry, when RF pulses are applied exactly in resonance with the nuclear spin transitions, we observe a slow decay in 20 $\mu$s. With a slight detuning of the RF frequency,  an additional oscillation is observed alongside the original decay. By fitting the oscillation using the equation $a\cdot$cos$(\omega \tau+\phi)e^{-\tau/T_{2,n}^*}+c$, we determine the inhomogeneous dephasing time to be $T_2^*$ = 16.6 $\mu$s and the oscillation frequency to be the same as the detuning. Figure \ref{fig:4}(d) shows the exponential decay measured by the nuclear spin Hahn Echo sequence, revealing a nuclear spin coherence time of $T_{2,n}$ = 162 $\mu$s. The nuclear spin  coherence time is comparable with the electron spin relaxation time $T_{1,e}$ = 144 $\mu$s (Supplemental Figure S15), suggesting the measurement is limited by the electronic spin lifetime.

\begin{figure}[!tph]
	\centering
	\includegraphics[width=0.5\textwidth]{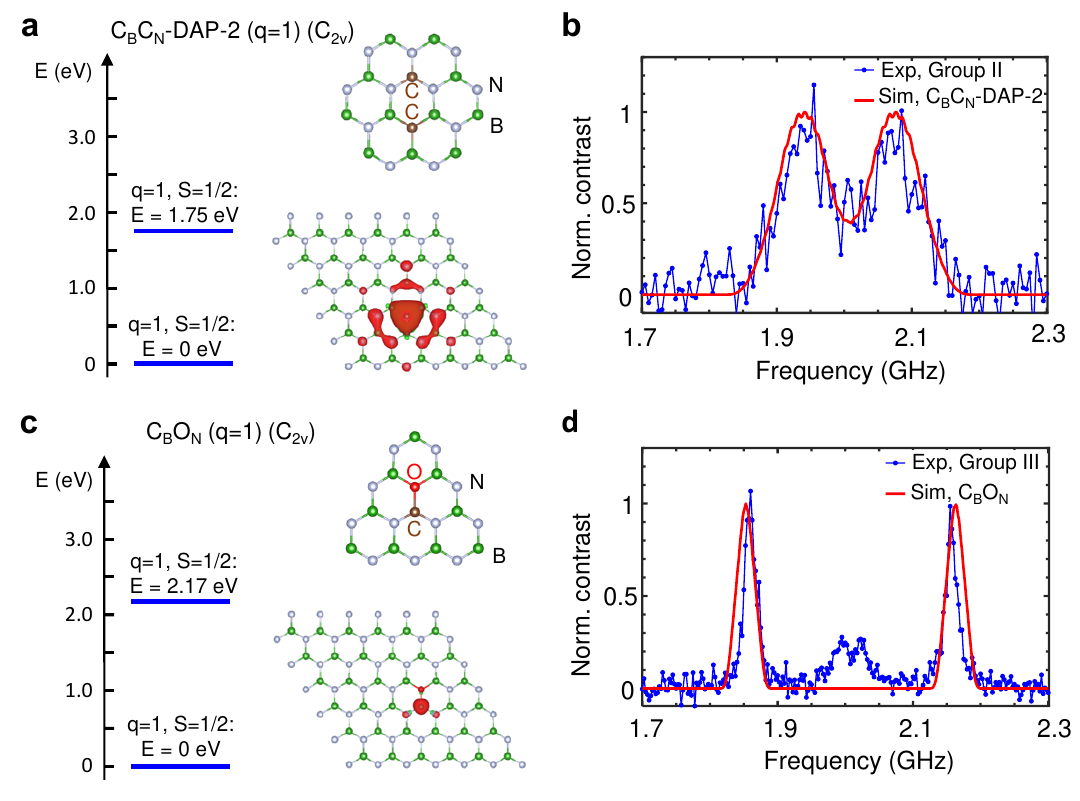}
	\caption{  \textbf{Candidates for the defect members in spin pairs.} (a) Chemical structure, spin density and energy-level diagram of the positively charged C$_B^+$C$_N^0$-DAP-2 defect. (b) Simulated ODMR spectrum of the positively charged C$_B^+$C$_N^0$-DAP-2 defect based on the calculated hyperfine coupling parameters. The simulation result ( red curve) is compared with the experimental result (blue curve). (c) Chemical structure, electronic wavefunction and energy level diagram of the positively charged C$_B$O$_N$ defect. (d) Simulated ODMR spectrum of the positively charged C$_B$O$_N$ defect based on the calculated hyperfine coupling parameters. The simulation result (red curve) is compared with the experimental result (blue curve). } \label{fig:5}
\end{figure}

\section*{Identifying chemical structures}

In addition to providing long-lived spin qubits, nuclear spin magnetic resonance offers valuable insight into identifying defect structures. Here, we perform first-principles calculations of the static properties and hyperfine interaction tensors to enable direct comparison with experimental data. Our calculations suggest that C$_B$C$_N$-DAP-$L$ \cite{auburger2021towards} is a likely member of the spin pair responsible for Group II defects (Figure \ref{fig:5}(a)). This dimer can exist in two charge states: C$_B^+$C$_N^0$-DAP-$L$ and  C$_B^0$C$_N^0$-DAP-$L$. The positively charged state  (C$_B^+$C$_N^0$-DAP-$L$) has an $S=1/2$ manifold, and our calculations predict a $^{13}$C$_N$ hyperfine coupling of 135~MHz (for $L=2$). This value closely matches the experimentally observed 130~MHz. The simulated ODMR spectrum aligns well with the measured ODMR, as shown in Figure \ref{fig:5}(b). In addition, the calculated hyperfine coupling for the nearest $^{11}$B nuclear spins is 17.6 MHz, resulting in an ODNMR resonance at 8.8 MHz, which closely matches the observed ODNMR resonance at 9.2 MHz in Figure \ref{fig:2}(h). Since the spin density is primarily localized at the C$_N$ site and is barely affected by the distance $L$, the hyperfine coupling strength remains nearly uncharged for different $L$ values (Supplemental Figure S32). 

To account for the observed 300~MHz hyperfine splitting in Group~III defects, we consider an oxygen substitution next to the $^{13}$C$_B$ site, forming a C$_B$O$_N$ defect \cite{guo2023coherent} as one member of the spin pair.  Oxygen impurities exist in hBN samples \cite{li2023prolonged} and can also be introduced during annealing due to residual oxygen in the vacuum chamber. The positively charged state (C$_B$O$_N^+$) of this defect hosts a spin-doublet manifold ($S=1/2$) (Figure~\ref{fig:5}(c)) with a hyperfine interaction strength $A_{zz} = 314$~MHz, 
in close agreement with our measured 300~MHz splitting. In Figure~\ref{fig:5}(d), the simulated ODMR linewidth induced by the nuclear spin bath is 23~MHz, comparable to the 16~MHz linewidth measured under weak microwave driving (Supplemental Figure~S13). The central peak (III-3) in Fig.~\ref{fig:5}(d)) is attributed to a second nearby spin defect. 

For both Group II and III defects, the structures discussed above correspond to the defect members in the spin pairs that exhibit strong hyperfine coupling. We expect a second, nearby donor or acceptor defect, couples to the proposed defect candidates to form a defect DAP complex.  The structure of the other defect remains unclear due to lack of characteristic hyperfine features. Further identification of both defect constituents may be possible by using high-purity hBN to suppress intrinsic $^{12}$C impurities, combined with higher $^{13}$C implantation density. This approach would improve the probability that both defects interact with $^{13}$C  nuclear spins, enabling more definitive structural assignments.

\section*{Conclusion}

In conclusion, we report the detection and coherent control of single $^{13}$C nuclear spins using single hBN spin defects at room temperature. We find three distinct defect groups in $^{13}$CO$_2$ implanted hBN samples, categorized based on their ODMR spectra. We observe both $S=1/2$ and $S=1$ spin states within a single hBN spin defect complex, which displays quantum coherence at room temperature and ODMR contrast up to 200$\%$. In addition, the electronic spin state can be readout using a reasonably long laser pulse of around 5 $\mu$s, yielding approximately one photon per readout pulse (Supplemental Figure S16) and a single-short spin readout efficiency of $\eta$ = 0.12 (see Methods).  

Leveraging the control of individual resonances within the well-resolved hyperfine structures, we demonstrate initialization, coherent control and readout of a single $^{13}$C nuclear spin using spin defects in Group II (Supplemental Section IV) and Group III. The nuclear spins exhibits coherence times that are orders of magnitudes longer than those of electronic spins in hBN, offering the potential for long-lived quantum registers. The well-resolved hyperfine structure, combined with the high readout efficiency of spin states enabled by the high ODMR contrasts and the extended nuclear spin coherence times, makes this approach promising for achieving single-shot readout of individual nuclear spins \cite{neumann2010single}. This capability is crucial for implementing quantum error correction protocols in a quantum register \cite{waldherr2014quantum}. Moreover, the  $^{13}$C nuclear spin can serve as a quantum memory to enhance quantum sensing with single hBN spin defects.



\begin{thebibliography}{37}
	\providecommand{\natexlab}[1]{#1}
	\providecommand{\url}[1]{\texttt{#1}}
	\expandafter\ifx\csname urlstyle\endcsname\relax
	\providecommand{\doi}[1]{doi: #1}\else
	\providecommand{\doi}{doi: \begingroup \urlstyle{rm}\Url}\fi
	
	\bibitem[Atat{\"u}re et~al.(2018)Atat{\"u}re, Englund, Vamivakas, Lee, and
	Wrachtrup]{atature2018material}
	Mete Atat{\"u}re, Dirk Englund, Nick Vamivakas, Sang-Yun Lee, and Joerg
	Wrachtrup.
	\newblock Material platforms for spin-based photonic quantum technologies.
	\newblock \emph{Nature Reviews Materials}, 3\penalty0 (5):\penalty0 38--51,
	2018.
	
	\bibitem[Wolfowicz et~al.(2021)Wolfowicz, Heremans, Anderson, Kanai, Seo, Gali,
	Galli, and Awschalom]{wolfowicz2021quantum}
	Gary Wolfowicz, F~Joseph Heremans, Christopher~P Anderson, Shun Kanai, Hosung
	Seo, Adam Gali, Giulia Galli, and David~D Awschalom.
	\newblock Quantum guidelines for solid-state spin defects.
	\newblock \emph{Nature Reviews Materials}, 6:\penalty0 906--925, 2021.
	
	\bibitem[Degen et~al.(2017)Degen, Reinhard, and Cappellaro]{degen2017quantum}
	Christian~L Degen, Friedemann Reinhard, and Paola Cappellaro.
	\newblock Quantum sensing.
	\newblock \emph{Reviews of modern physics}, 89:\penalty0 035002, 2017.
	
	\bibitem[Du et~al.(2024)Du, Shi, Kong, Jelezko, and Wrachtrup]{du2024single}
	Jiangfeng Du, Fazhan Shi, Xi~Kong, Fedor Jelezko, and J{\"o}rg Wrachtrup.
	\newblock Single-molecule scale magnetic resonance spectroscopy using quantum
	diamond sensors.
	\newblock \emph{Reviews of Modern Physics}, 96:\penalty0 025001, 2024.
	
	\bibitem[Pompili et~al.(2021)Pompili, Hermans, Baier, Beukers, Humphreys,
	Schouten, Vermeulen, Tiggelman, dos Santos~Martins, Dirkse,
	et~al.]{pompili2021realization}
	Matteo Pompili, Sophie~LN Hermans, Simon Baier, Hans~KC Beukers, Peter~C
	Humphreys, Raymond~N Schouten, Raymond~FL Vermeulen, Marijn~J Tiggelman,
	Laura dos Santos~Martins, Bas Dirkse, et~al.
	\newblock Realization of a multinode quantum network of remote solid-state
	qubits.
	\newblock \emph{Science}, 372:\penalty0 259--264, 2021.
	
	\bibitem[Knaut et~al.(2024)Knaut, Suleymanzade, Wei, Assumpcao, Stas, Huan,
	Machielse, Knall, Sutula, Baranes, et~al.]{knaut2024entanglement}
	CM~Knaut, A~Suleymanzade, Y-C Wei, DR~Assumpcao, P-J Stas, YQ~Huan,
	B~Machielse, EN~Knall, M~Sutula, G~Baranes, et~al.
	\newblock Entanglement of nanophotonic quantum memory nodes in a telecom
	network.
	\newblock \emph{Nature}, 629:\penalty0 573--578, 2024.
	
	\bibitem[Chejanovsky et~al.(2021)Chejanovsky, Mukherjee, Geng, Chen, Kim,
	Denisenko, Finkler, Taniguchi, Watanabe, Dasari,
	et~al.]{chejanovsky2021single}
	Nathan Chejanovsky, Amlan Mukherjee, Jianpei Geng, Yu-Chen Chen, Youngwook Kim,
	Andrej Denisenko, Amit Finkler, Takashi Taniguchi, Kenji Watanabe, Durga
	Bhaktavatsala~Rao Dasari, et~al.
	\newblock Single-spin resonance in a van der {Waals} embedded paramagnetic
	defect.
	\newblock \emph{Nature materials}, 20:\penalty0 1079--1084, 2021.
	
	\bibitem[Mendelson et~al.(2021)Mendelson, Chugh, Reimers, Cheng, Gottscholl,
	Long, Mellor, Zettl, Dyakonov, Beton, et~al.]{mendelson2021identifying}
	Noah Mendelson, Dipankar Chugh, Jeffrey~R Reimers, Tin~S Cheng, Andreas
	Gottscholl, Hu~Long, Christopher~J Mellor, Alex Zettl, Vladimir Dyakonov,
	Peter~H Beton, et~al.
	\newblock Identifying carbon as the source of visible single-photon emission
	from hexagonal boron nitride.
	\newblock \emph{Nature materials}, 20:\penalty0 321--328, 2021.
	
	\bibitem[Stern et~al.(2022)Stern, Gu, Jarman, Eizagirre~Barker, Mendelson,
	Chugh, Schott, Tan, Sirringhaus, Aharonovich, et~al.]{stern2022room}
	Hannah~L Stern, Qiushi Gu, John Jarman, Simone Eizagirre~Barker, Noah
	Mendelson, Dipankar Chugh, Sam Schott, Hoe~H Tan, Henning Sirringhaus, Igor
	Aharonovich, et~al.
	\newblock Room-temperature optically detected magnetic resonance of single
	defects in hexagonal boron nitride.
	\newblock \emph{Nature communications}, 13\penalty0 (1):\penalty0 618, 2022.
	
	\bibitem[Guo et~al.(2023)Guo, Li, Liu, Yang, Zeng, Yu, Meng, Li, Wang, Xie,
	et~al.]{guo2023coherent}
	Nai-Jie Guo, Song Li, Wei Liu, Yuan-Ze Yang, Xiao-Dong Zeng, Shang Yu, Yu~Meng,
	Zhi-Peng Li, Zhao-An Wang, Lin-Ke Xie, et~al.
	\newblock Coherent control of an ultrabright single spin in hexagonal boron
	nitride at room temperature.
	\newblock \emph{Nature Communications}, 14\penalty0 (1):\penalty0 2893, 2023.
	
	\bibitem[Stern et~al.(2024)Stern, M.~Gilardoni, Gu, Eizagirre~Barker, Powell,
	Deng, Fraser, Follet, Li, Ramsay, et~al.]{stern2023quantum}
	Hannah~L Stern, Carmem M.~Gilardoni, Qiushi Gu, Simone Eizagirre~Barker,
	Oliver~FJ Powell, Xiaoxi Deng, Stephanie~A Fraser, Louis Follet, Chi Li,
	Andrew~J Ramsay, et~al.
	\newblock A quantum coherent spin in hexagonal boron nitride at ambient
	conditions.
	\newblock \emph{Nature Materials}, 23\penalty0 (10):\penalty0 1379--1385, 2024.
	
	\bibitem[Shen et~al.(2024)Shen, Xiao, and Cao]{shen2024proximity}
	Lingnan Shen, Di~Xiao, and Ting Cao.
	\newblock Proximity-induced exchange interaction: A new pathway for quantum
	sensing using spin centers in hexagonal boron nitride.
	\newblock \emph{The Journal of Physical Chemistry Letters}, 15\penalty0
	(16):\penalty0 4359--4366, 2024.
	
	\bibitem[Gottscholl et~al.(2020)Gottscholl, Kianinia, Soltamov, Orlinskii,
	Mamin, Bradac, Kasper, Krambrock, Sperlich, Toth,
	et~al.]{gottscholl2020initialization}
	Andreas Gottscholl, Mehran Kianinia, Victor Soltamov, Sergei Orlinskii, Georgy
	Mamin, Carlo Bradac, Christian Kasper, Klaus Krambrock, Andreas Sperlich,
	Milos Toth, et~al.
	\newblock Initialization and read-out of intrinsic spin defects in a van der
	{Waals} crystal at room temperature.
	\newblock \emph{Nature materials}, 19:\penalty0 540--545, 2020.
	
	\bibitem[Caldwell et~al.(2019)Caldwell, Aharonovich, Cassabois, Edgar, Gil, and
	Basov]{caldwell2019photonics}
	Joshua~D Caldwell, Igor Aharonovich, Guillaume Cassabois, James~H Edgar,
	Bernard Gil, and DN~Basov.
	\newblock Photonics with hexagonal boron nitride.
	\newblock \emph{Nature Reviews Materials}, 4:\penalty0 552--567, 2019.
	
	\bibitem[Gottscholl et~al.(2021{\natexlab{a}})Gottscholl, Diez, Soltamov,
	Kasper, Krau{\ss}e, Sperlich, Kianinia, Bradac, Aharonovich, and
	Dyakonov]{gottscholl2021spin}
	Andreas Gottscholl, Matthias Diez, Victor Soltamov, Christian Kasper, Dominik
	Krau{\ss}e, Andreas Sperlich, Mehran Kianinia, Carlo Bradac, Igor
	Aharonovich, and Vladimir Dyakonov.
	\newblock Spin defects in {hBN} as promising temperature, pressure and magnetic
	field quantum sensors.
	\newblock \emph{Nature communications}, 12:\penalty0 4480, 2021{\natexlab{a}}.
	
	\bibitem[Healey et~al.(2023)Healey, Scholten, Yang, Scott, Abrahams, Robertson,
	Hou, Guo, Rahman, Lu, et~al.]{healey2022quantum}
	AJ~Healey, SC~Scholten, T~Yang, JA~Scott, GJ~Abrahams, IO~Robertson, XF~Hou,
	YF~Guo, S~Rahman, Y~Lu, et~al.
	\newblock Quantum microscopy with van der {Waals} heterostructures.
	\newblock \emph{Nature Physics}, 19:\penalty0 87–91, 2023.
	
	\bibitem[Huang et~al.(2022)Huang, Zhou, Chen, Lu, McLaughlin, Li, Alghamdi,
	Djugba, Shi, Wang, et~al.]{huang2022wide}
	Mengqi Huang, Jingcheng Zhou, Di~Chen, Hanyi Lu, Nathan~J McLaughlin, Senlei
	Li, Mohammed Alghamdi, Dziga Djugba, Jing Shi, Hailong Wang, et~al.
	\newblock Wide field imaging of van der {Waals} ferromagnet {Fe$_3$GeTe$_2$} by
	spin defects in hexagonal boron nitride.
	\newblock \emph{Nature communications}, 13:\penalty0 5369, 2022.
	
	\bibitem[Lyu et~al.(2022)Lyu, Tan, Wu, Zhang, Zhang, Mu,
	Z{\'u}{\~n}iga-P{\'e}rez, Cai, and Gao]{lyu2022strain}
	Xiaodan Lyu, Qinghai Tan, Lishu Wu, Chusheng Zhang, Zhaowei Zhang, Zhao Mu,
	Jes{\'u}s Z{\'u}{\~n}iga-P{\'e}rez, Hongbing Cai, and Weibo Gao.
	\newblock Strain quantum sensing with spin defects in hexagonal boron nitride.
	\newblock \emph{Nano Letters}, 22:\penalty0 6553--6559, 2022.
	
	\bibitem[Vaidya et~al.(2023)Vaidya, Gao, Dikshit, Aharonovich, and
	Li]{vaidya2023quantum}
	Sumukh Vaidya, Xingyu Gao, Saakshi Dikshit, Igor Aharonovich, and Tongcang Li.
	\newblock Quantum sensing and imaging with spin defects in hexagonal boron
	nitride.
	\newblock \emph{Advances in Physics: X}, 8:\penalty0 2206049, 2023.
	
	\bibitem[Gottscholl et~al.(2021{\natexlab{b}})Gottscholl, Diez, Soltamov,
	Kasper, Sperlich, Kianinia, Bradac, Aharonovich, and
	Dyakonov]{gottscholl2021room}
	Andreas Gottscholl, Matthias Diez, Victor Soltamov, Christian Kasper, Andreas
	Sperlich, Mehran Kianinia, Carlo Bradac, Igor Aharonovich, and Vladimir
	Dyakonov.
	\newblock Room temperature coherent control of spin defects in hexagonal boron
	nitride.
	\newblock \emph{Science Advances}, 7:\penalty0 eabf3630, 2021{\natexlab{b}}.
	
	\bibitem[Gong et~al.(2023)Gong, He, Gao, Ju, Liu, Ye, Henriksen, Li, and
	Zu]{gong2023coherent}
	Ruotian Gong, Guanghui He, Xingyu Gao, Peng Ju, Zhongyuan Liu, Bingtian Ye,
	Erik~A Henriksen, Tongcang Li, and Chong Zu.
	\newblock Coherent dynamics of strongly interacting electronic spin defects in
	hexagonal boron nitride.
	\newblock \emph{Nature Communications}, 14\penalty0 (1):\penalty0 3299, 2023.
	
	\bibitem[Rizzato et~al.(2023)Rizzato, Schalk, Mohr, Hermann, Leibold,
	Bruckmaier, Salvitti, Qian, Ji, Astakhov, et~al.]{rizzato2023extending}
	Roberto Rizzato, Martin Schalk, Stephan Mohr, Jens~C Hermann, Joachim~P
	Leibold, Fleming Bruckmaier, Giovanna Salvitti, Chenjiang Qian, Peirui Ji,
	Georgy~V Astakhov, et~al.
	\newblock Extending the coherence of spin defects in {hBN} enables advanced
	qubit control and quantum sensing.
	\newblock \emph{Nature Communications}, 14\penalty0 (1):\penalty0 5089, 2023.
	
	\bibitem[Dutt et~al.(2007)Dutt, Childress, Jiang, Togan, Maze, Jelezko, Zibrov,
	Hemmer, and Lukin]{dutt2007quantum}
	MV~Gurudev Dutt, L~Childress, L~Jiang, E~Togan, J~Maze, F~Jelezko, AS~Zibrov,
	PR~Hemmer, and MD~Lukin.
	\newblock Quantum register based on individual electronic and nuclear spin
	qubits in diamond.
	\newblock \emph{Science}, 316:\penalty0 1312--1316, 2007.
	
	\bibitem[Zaiser et~al.(2016)Zaiser, Rendler, Jakobi, Wolf, Lee, Wagner,
	Bergholm, Schulte-Herbr{\"u}ggen, Neumann, and
	Wrachtrup]{zaiser2016enhancing}
	Sebastian Zaiser, Torsten Rendler, Ingmar Jakobi, Thomas Wolf, Sang-Yun Lee,
	Samuel Wagner, Ville Bergholm, Thomas Schulte-Herbr{\"u}ggen, Philipp
	Neumann, and J{\"o}rg Wrachtrup.
	\newblock Enhancing quantum sensing sensitivity by a quantum memory.
	\newblock \emph{Nature communications}, 7\penalty0 (1):\penalty0 12279, 2016.
	
	\bibitem[Aslam et~al.(2017)Aslam, Pfender, Neumann, Reuter, Zappe,
	F{\'a}varo~de Oliveira, Denisenko, Sumiya, Onoda, Isoya,
	et~al.]{aslam2017nanoscale}
	Nabeel Aslam, Matthias Pfender, Philipp Neumann, Rolf Reuter, Andrea Zappe,
	Felipe F{\'a}varo~de Oliveira, Andrej Denisenko, Hitoshi Sumiya, Shinobu
	Onoda, Junichi Isoya, et~al.
	\newblock Nanoscale nuclear magnetic resonance with chemical resolution.
	\newblock \emph{Science}, 357\penalty0 (6346):\penalty0 67--71, 2017.
	
	\bibitem[Neumann et~al.(2010)Neumann, Beck, Steiner, Rempp, Fedder, Hemmer,
	Wrachtrup, and Jelezko]{neumann2010single}
	Philipp Neumann, Johannes Beck, Matthias Steiner, Florian Rempp, Helmut Fedder,
	Philip~R Hemmer, J{\"o}rg Wrachtrup, and Fedor Jelezko.
	\newblock Single-shot readout of a single nuclear spin.
	\newblock \emph{Science}, 329:\penalty0 542--544, 2010.
	
	\bibitem[Waldherr et~al.(2014)Waldherr, Wang, Zaiser, Jamali,
	Schulte-Herbr{\"u}ggen, Abe, Ohshima, Isoya, Du, Neumann,
	et~al.]{waldherr2014quantum}
	Gerald Waldherr, Yiqing Wang, S~Zaiser, M~Jamali, T~Schulte-Herbr{\"u}ggen,
	H~Abe, T~Ohshima, J~Isoya, JF~Du, P~Neumann, et~al.
	\newblock Quantum error correction in a solid-state hybrid spin register.
	\newblock \emph{Nature}, 506:\penalty0 204--207, 2014.
	
	\bibitem[Gao et~al.(2022)Gao, Vaidya, Li, Ju, Jiang, Xu, Allcca, Shen,
	Taniguchi, Watanabe, et~al.]{gao2022nuclear}
	Xingyu Gao, Sumukh Vaidya, Kejun Li, Peng Ju, Boyang Jiang, Zhujing Xu, Andres
	E~Llacsahuanga Allcca, Kunhong Shen, Takashi Taniguchi, Kenji Watanabe,
	et~al.
	\newblock Nuclear spin polarization and control in hexagonal boron nitride.
	\newblock \emph{Nature Materials}, 21:\penalty0 1024--1028, 2022.
	
	\bibitem[Gong et~al.(2024)Gong, Du, Janzen, Liu, Liu, He, Ye, Li, Yao, Edgar,
	et~al.]{gong2024isotope}
	Ruotian Gong, Xinyi Du, Eli Janzen, Vincent Liu, Zhongyuan Liu, Guanghui He,
	Bingtian Ye, Tongcang Li, Norman~Y Yao, James~H Edgar, et~al.
	\newblock Isotope engineering for spin defects in van der waals materials.
	\newblock \emph{Nature Communications}, 15:\penalty0 104, 2024.
	
	\bibitem[Clua-Provost et~al.(2023)Clua-Provost, Durand, Mu, Rastoin,
	Frauni{\'e}, Janzen, Schutte, Edgar, Seine, Claverie,
	et~al.]{clua2023isotopic}
	T~Clua-Provost, A~Durand, Z~Mu, T~Rastoin, J~Frauni{\'e}, E~Janzen, H~Schutte,
	JH~Edgar, G~Seine, A~Claverie, et~al.
	\newblock Isotopic control of the boron-vacancy spin defect in hexagonal boron
	nitride.
	\newblock \emph{Physical Review Letters}, 131:\penalty0 126901, 2023.
	
	\bibitem[Ru et~al.(2024)Ru, Jiang, Liang, Kenny, Cai, Lyu, Cernansky, Zhou,
	Yang, Watanabe, et~al.]{ru2024robust}
	Shihao Ru, Zhengzhi Jiang, Haidong Liang, Jonathan Kenny, Hongbing Cai, Xiaodan
	Lyu, Robert Cernansky, Feifei Zhou, Yuzhe Yang, Kenji Watanabe, et~al.
	\newblock Robust nuclear spin polarization via ground-state level anti-crossing
	of boron vacancy defects in hexagonal boron nitride.
	\newblock \emph{Physical Review Letters}, 132:\penalty0 266801, 2024.
	

	\bibitem[Robertson et~al.(2024)Robertson, Whitefield, Scholten, Singh, Healey,
	Reineck, Kianinia, Barcza, Ivády, Broadway, Aharonovich, and
	Tetienne]{robertson2024universal}
	Islay~O Robertson, Benjamin Whitefield, Sam~C Scholten, Priya Singh,
	Alexander~J Healey, Philipp Reineck, Mehran Kianinia, Gergely Barcza, Viktor
	Ivády, David~A Broadway, Igor Aharonovich, and Jean-Philippe Tetienne.
	\newblock A charge transfer mechanism for optically addressable solid-state
	spin pairs.
	\newblock \emph{arXiv preprint arXiv:2407.13148}, 2024.
	
	\bibitem[Patel et~al.(2024)Patel, Fishman, Huang, Gusdorff, Fehr, Hopper,
	Breitweiser, Porat, Flatt{\'e}, and Bassett]{patel2024room}
	Raj~N Patel, Rebecca~EK Fishman, Tzu-Yung Huang, Jordan~A Gusdorff, David~A
	Fehr, David~A Hopper, S~Alex Breitweiser, Benjamin Porat, Michael~E
	Flatt{\'e}, and Lee~C Bassett.
	\newblock Room temperature dynamics of an optically addressable single spin in
	hexagonal boron nitride.
	\newblock \emph{Nano Letters}, 24:\penalty0 7623--7628, 2024.
	
	\bibitem[Tan et~al.(2022)Tan, Lai, Liu, Guo, Xue, Dou, Sun, Deng, Tan,
	Aharonovich, et~al.]{tan2022donor}
	Qinghai Tan, Jia-Min Lai, Xue-Lu Liu, Dan Guo, Yongzhou Xue, Xiuming Dou,
	Bao-Quan Sun, Hui-Xiong Deng, Ping-Heng Tan, Igor Aharonovich, et~al.
	\newblock Donor--acceptor pair quantum emitters in hexagonal boron nitride.
	\newblock \emph{Nano Letters}, 22:\penalty0 1331--1337, 2022.
	
	\bibitem[Pelliciari et~al.(2024)Pelliciari, Mejia, Woods, Gu, Li, Chand, Fan,
	Watanabe, Taniguchi, Bisogni, et~al.]{pelliciari2024elementary}
	Jonathan Pelliciari, Enrique Mejia, John~M Woods, Yanhong Gu, Jiemin Li,
	Saroj~B Chand, Shiyu Fan, Kenji Watanabe, Takashi Taniguchi, Valentina
	Bisogni, et~al.
	\newblock Elementary excitations of single-photon emitters in hexagonal boron
	nitride.
	\newblock \emph{Nature Materials}, 23:\penalty0 1230--1236, 2024.
	
	\bibitem[Auburger and Gali(2021)]{auburger2021towards}
	Philipp Auburger and Adam Gali.
	\newblock Towards {\it ab initio} identification of paramagnetic substitutional
	carbon defects in hexagonal boron nitride acting as quantum bits.
	\newblock \emph{Physical Review B}, 104:\penalty0 075410, 2021.
	
	\bibitem[Li et~al.(2023)Li, Ichihara, Park, He, Kozawa, Wen, Koman, Zeng,
	Kuehne, Yuan, et~al.]{li2023prolonged}
	Sylvia~Xin Li, Takeo Ichihara, Hyoju Park, Guangwei He, Daichi Kozawa, Yi~Wen,
	Volodymyr~B Koman, Yuwen Zeng, Matthias Kuehne, Zhe Yuan, et~al.
	\newblock Prolonged photostability in hexagonal boron nitride quantum emitters.
	\newblock \emph{Communications Materials}, 4:\penalty0 19, 2023.
	
\end{thebibliography}

\begin{thebibliography}{13}
	
	\makeatletter
	\addtocounter{NAT@ctr}{37}
	\makeatother
	
	\expandafter\ifx\csname url\endcsname\relax
	\def\url#1{\texttt{#1}}\fi
	\expandafter\ifx\csname urlprefix\endcsname\relax\def\urlprefix{URL }\fi
	\providecommand{\bibinfo}[2]{#2}
	\providecommand{\eprint}[2][]{\url{#2}}
	
	\bibitem[Hopper et~al.(2018)Hopper, Shulevitz, and Bassett]{hopper2018spin}
	David~A Hopper, Henry~J Shulevitz, and Lee~C Bassett.
	\newblock Spin readout techniques of the nitrogen-vacancy center in diamond.
	\newblock \emph{Micromachines}, 9:\penalty0 437, 2018.
	
	\bibitem[Debroux et~al.(2021)Debroux, Michaels, Purser, Wan, Trusheim,
	Arjona~Mart{\'\i}nez, Parker, Stramma, Chen, De~Santis,
	et~al.]{debroux2021quantum}
	Romain Debroux, Cathryn~P Michaels, Carola~M Purser, Noel Wan, Matthew~E
	Trusheim, Jes{\'u}s Arjona~Mart{\'\i}nez, Ryan~A Parker, Alexander~M Stramma,
	Kevin~C Chen, Lorenzo De~Santis, et~al.
	\newblock Quantum control of the tin-vacancy spin qubit in diamond.
	\newblock \emph{Physical Review X}, 11:\penalty0 041041, 2021.
	
	\bibitem[Giannozzi et~al.(2009)Giannozzi, Baroni, Bonini, Calandra, Car,
	Cavazzoni, Ceresoli, Chiarotti, Cococcioni, Dabo, {Dal Corso}, de~Gironcoli,
	Fabris, Fratesi, Gebauer, Gerstmann, Gougoussis, Kokalj, Lazzeri,
	Martin-Samos, Marzari, Mauri, Mazzarello, Paolini, Pasquarello, Paulatto,
	Sbraccia, Scandolo, Sclauzero, Seitsonen, Smogunov, Umari, and
	Wentzcovitch]{QE}
	Paolo Giannozzi, Stefano Baroni, Nicola Bonini, Matteo Calandra, Roberto Car,
	Carlo Cavazzoni, Davide Ceresoli, Guido~L Chiarotti, Matteo Cococcioni,
	Ismaila Dabo, Andrea {Dal Corso}, Stefano de~Gironcoli, Stefano Fabris, Guido
	Fratesi, Ralph Gebauer, Uwe Gerstmann, Christos Gougoussis, Anton Kokalj,
	Michele Lazzeri, Layla Martin-Samos, Nicola Marzari, Francesco Mauri,
	Riccardo Mazzarello, Stefano Paolini, Alfredo Pasquarello, Lorenzo Paulatto,
	Carlo Sbraccia, Sandro Scandolo, Gabriele Sclauzero, Ari~P Seitsonen,
	Alexander Smogunov, Paolo Umari, and Renata~M Wentzcovitch.
	\newblock {QUANTUM ESPRESSO: A Modular and Open-Source Software Project for
		Quantum Simulations of Materials.}
	\newblock \emph{J. Phys.: Condens. Matter}, 21:\penalty0 395502, sep 2009.
	
	\bibitem[Heyd et~al.(2003)Heyd, Scuseria, and Ernzerhof]{heyd2003hybrid}
	Jochen Heyd, Gustavo~E Scuseria, and Matthias Ernzerhof.
	\newblock {Hybrid Functionals Based on a Screened Coulomb Potential}.
	\newblock \emph{J. Chem. Phys.}, 118:\penalty0 8207--8215, 2003.
	\newblock \doi{10.1063/1.1564060}.
	
	\bibitem[Heyd et~al.(2006)Heyd, Scuseria, and Ernzerhof]{HSE06}
	Jochen Heyd, Gustavo~E. Scuseria, and Matthias Ernzerhof.
	\newblock {Erratum: “Hybrid functionals based on a screened Coulomb
		potential” [J. Chem. Phys. 118, 8207 (2003)]}.
	\newblock \emph{The Journal of Chemical Physics}, 124:\penalty0 219906, 06
	2006.
	\newblock ISSN 0021-9606.
	\newblock \doi{10.1063/1.2204597}.
	
	\bibitem[Hamann(2013)]{ONCV1}
	D.~R. Hamann.
	\newblock {Optimized Norm-Conserving Vanderbilt Pseudopotentials}.
	\newblock \emph{Phys. Rev. B}, 88:\penalty0 085117, aug 2013.
	
	\bibitem[Schlipf and Gygi(2015)]{ONCV2}
	Martin Schlipf and Fran{\c{c}}ois Gygi.
	\newblock {Optimization Algorithm for the Generation of ONCV Pseudopotentials}.
	\newblock \emph{Comput. Phys. Commun.}, 196:\penalty0 36--44, 2015.
	
	\bibitem[Tantardini et~al.(2022)Tantardini, Kvashnin, and
	Ceresoli]{tantardini2022gipaw}
	Christian Tantardini, Alexander~G Kvashnin, and Davide Ceresoli.
	\newblock Gipaw pseudopotentials of d elements for solid-state nmr.
	\newblock \emph{Mater.}, 15:\penalty0 3347, 2022.
	
	\bibitem[Smart et~al.(2021)Smart, Li, Xu, and Ping]{smart2021intersystem}
	Tyler~J Smart, Kejun Li, Junqing Xu, and Yuan Ping.
	\newblock {Intersystem Crossing and Exciton--Defect Coupling of Spin Defects in
		Hexagonal Boron Nitride}.
	\newblock \emph{npj Comput. Mater.}, 7:\penalty0 1--8, 2021.
	\newblock \doi{10.1038/s41524-021-00525-5}.
	
	\bibitem[Mackoit-Sinkevi{\v{c}}ien{\.e}
	et~al.(2019)Mackoit-Sinkevi{\v{c}}ien{\.e}, Maciaszek, Van~de Walle, and
	Alkauskas]{mackoit2019carbon}
	M~Mackoit-Sinkevi{\v{c}}ien{\.e}, Marek Maciaszek, Chris~G Van~de Walle, and
	Audrius Alkauskas.
	\newblock {Carbon Dimer Defect as a Source of the 4.1 eV Luminescence in
		Hexagonal Boron Nitride}.
	\newblock \emph{Appl. Phys. Lett.}, 115:\penalty0 212101, 2019.
	
	\bibitem[Varini et~al.(2013)Varini, Ceresoli, Martin-Samos, Girotto, and
	Cavazzoni]{varini2013enhancement}
	Nicola Varini, Davide Ceresoli, Layla Martin-Samos, Ivan Girotto, and Carlo
	Cavazzoni.
	\newblock Enhancement of dft-calculations at petascale: nuclear magnetic
	resonance, hybrid density functional theory and car--parrinello calculations.
	\newblock \emph{Comput. Phys. Commun.}, 184:\penalty0 1827--1833, 2013.
	
	\bibitem[Ma et~al.(2020)Ma, Govoni, and Galli]{ma2020pyzfs}
	He~Ma, Marco Govoni, and Giulia Galli.
	\newblock {PyZFS: A Python Package for First-Principles Calculations of
		Zero-Field Splitting Tensors}.
	\newblock \emph{J. Open Source Softw.}, 5:\penalty0 2160, 2020.
	\newblock \doi{10.21105/joss.02160}.
	
	\bibitem[Stoll and Schweiger(2006)]{stoll2006easyspin}
	Stefan Stoll and Arthur Schweiger.
	\newblock Easyspin, a comprehensive software package for spectral simulation
	and analysis in epr.
	\newblock \emph{Journal of magnetic resonance}, 178:\penalty0 42--55, 2006.
	
\end{thebibliography}

\clearpage 
\newpage

\section* {Methods}

\textbf{Sample preparation}

The hBN thin flakes were tape-exfoliated from a monocrystalline hBN crystal and transferred onto Si/SiO$_2$ substrates. Then we irradiated the hBN flakes  with 2.5 keV  $^{13}$CO$_2$ (99.0$\%$ $^{13}$C, Sigma-Aldrich) ions with a dose density of 10$^{12}$ cm$^{-2}$ using a home-built ion implanter. The sample is then annealed at 1000 $^\circ$C at 10$^{-5}$ torr for 2 hours to activate the carbon-related defects. For ODMR measurements, we transferred the hBN flakes to a coplanar waveguide using the standard dry transfer method with propylene carbonate stamps. The waveguide is made of 200 nm thick silver with a 4 nm thick Al$_2$O$_3$ layer on top.

\vspace{0.1in}
\textbf{Sensitivity of a single spin defect}

The ODMR contrast varies between defects and can reach as high as 200$\%$ (Supplemental Figure S24). Among the over 100 spin defects investigated, approximately 25 $\%$ exhibit a contrast higher than  10 $\%$. A single hBN spin defect in our sample has a typical sensitivity of 5~$\mu$T/$\sqrt{\rm Hz}$ for DC magnetic field sensing, calculated using  $(8\pi/3\sqrt{3})\cdot(1/\gamma_e)\cdot(\Delta\nu/C\sqrt{I})$\cite{gong2024isotope}, where $\Delta\nu$ is linewidth (20 MHz), $C$ is the contrast (30 $\%$) and $I$ is the photon count rate (170 kcts/s). Additionally, while most Group II and Group III defects exhibit stable behaviors under a weak laser excitation ($\le$ 15 $\mu$W), the stability of Group I defects varies significantly.

\vspace{0.1in}
\textbf{Estimation of nuclear spin polarization}

We estimate the polarization of the $^{13}$C nuclear spin by evaluating the imbalance between III-2 and III-4 in the ODMR spectrum. The ODMR is taken after the SWAP gate to transfer the electron polarization to the $^{13}$C nuclear spin. By using the fitted relative populations of the hyperfine basis states, the polarization can be calculated by the equation
\begin{equation}
	P = \frac{\sum_{m_I} m_I\rho_{m_I}}{I\sum_{m_I}\rho_{m_I}}=\frac{ \rho_{1/2}-\rho_{-1/2}}{\rho_{1/2}+\rho_{-1/2}}.
\end{equation}

\vspace{0.1in}
\textbf{Spin readout efficiency}

The efficiency of a single-shot spin readout is an important factor to estimate how efficiently one can determine the electronic spin state of a spin defect, which is highly dependent on the defect properties.  The readout efficiency is defined by the signal-to-noise ratio from a single readout pulse and can be express as \cite{hopper2018spin}
\begin{equation}
	\eta_s = 1/\sigma_{s}=\left(1+2\frac{\alpha_0+\alpha_1}{(\alpha_0-\alpha_1)^2}\right)^{-1/2}.
\end{equation}
where $\alpha_0$ and $\alpha_1$ are the mean numbers of detected photons for a single measurement of the brighter state and darker state, respectively. We estimate the efficiency based on the pulsed ODMR measurements. The pulsed ODMR contrast is 17.5$\%$ when we set readout duration at 5 $\mu$s under a P$_{MW}$=60 mW microwave drive. The contrast reaches 28$\%$ when P$_{MW}$=2 W. For each readout laser pulse, we obtain approximately 0.9 photon from the darker state under the 15 $\mu$W laser pumping. These yield the efficiency of 0.08 and 0.12 for P$_{MW}$=60 mW and P$_{MW}$=2 W, respectively. See details in Supplemental Section IV.

\vspace{0.1in}
\textbf{Gate fidelity}

For Rabi oscillations limited by a pure dephasing process, we can write the $\pi$-gate fidelity as $F_{\pi}=0.5(1+exp(-1/Q_\pi))$, where $Q_\pi$ = $T_{Rabi}/T_{\pi}$ is the quality factor of a $\pi$ gate\cite{debroux2021quantum}. We extract the coherence time and $\pi$-gate time of nuclear spin Rabi by fitting the results in Figure 3 to the function $C(\tau ) = a \cdot\sin \left( {\pi \tau /{T_\pi } + b} \right) \cdot \exp \left( -{\tau /{T_{Rabi}}} \right) + d $, where $C(\tau)$ is the signal contrast of Rabi. As a results, we obtain a $F_{\pi,n}$ = 99.75$\%$ $\pi$-gate fidelity, with $T_{\pi,n}$ = 0.60 $\mu$s and $T_{Rabi,n}$ = 117 $\mu$s. Similarly, we also estimate the electronic spin $\pi$-gate fidelity to be 96.2$\%$, using the same defect and transition in the nuclear spin control experiments.

\vspace{0.1in}
\textbf{Density-functional theory calculations}

We use Quantum Espresso (QE)~\cite{QE}, an open-source plane-wave software, to perform the density-function theory (DFT) calculation. Both the Perdew-Burke-Ernzerhof (PBE) functional and the Heyd-Scuseria-Ernzerhof (HSE) hybrid functional (the factor of 0.32 for Fock 
exchange)~\cite{heyd2003hybrid, HSE06} are employed for the exchange-correlation interaction. We use Optimized Norm-Conserving Vanderbilt (ONCV) pseudopotential~\cite{ONCV1,ONCV2} for the calculations of excitation energy, and the GIPAW pseudopotential~\cite{tantardini2022gipaw} for the calculation of hyperfine interaction parameters and zero-field splitting (ZFS). We set the kinetic energy cutoff to be 55 Ry, which is adequate for converging the relevant properties. Geometry optimizations are carried out with a force threshold of 
0.001 Ry$/$Bohr.  We select the $6\times6\times1$ or higher supercell size of hBN for the calculations of hyperfine parameter and excitation energies. For these calculations, we sample a k-point mesh of $3\times3\times1$ for the calculation of excitation energies~\cite{smart2021intersystem}, and $\Gamma$ point for the hyperfine parameters and ZFS~\cite{gao2022nuclear,smart2021intersystem}. We calculate the zero-phonon line (ZPL) by the constraint occupation DFT (CDFT) method~\cite{mackoit2019carbon}, the hyperfine parameters using the QE-GIPAW code~\cite{varini2013enhancement}, the ZFS by using the ZFS code~\cite{smart2021intersystem}, and we cross compare results between ZFS code and 
the PyZFS code~\cite{ma2020pyzfs}.   The key results are summarized in Supplemental Table S2-S3.

\vspace{0.1in}
\textbf{Simulation of ODMR spectrum}

The continuous wave (cw) ODMR spectra shown in Figure \ref{fig:5} are simulated using the MATLAB toolbox EASYSPIN \cite{stoll2006easyspin} based on data from the {\it ab initio} calculations.  EASYSPIN also takes the nuclear Zeeman and
quadrupole interaction into account. Therefore, the cw ODMR linewidth can be determined according to the hyperfine couplings with the most abundant nuclear-spin-active isotopes: $^{13}$C,$^{11}$B and $^{14}$N. In our simulation, we consider a $^{13}$C nuclear spin, ten nearest $^{11}$B nuclear spins and two proximate $^{14}$N nuclear spins. The other nuclei, located further away, couple more weakly to the electron, scaling with $\propto$ $1/r^3$ (where $r$ is the distance from the central carbon site), and thereby have a negligible effect on the ODMR linewidth.

\vspace{0.1in}
\textbf{Spin pair model}

The S = 1 transitions are consistently observed alongside the S = 1/2 transitions within the same emitters. To explain the coexistence of both spin transitions in ODMR, we employ a weakly coupled spin-pair model \cite{robertson2024universal, patel2024room} and perform numerical simulations to investigate the underlying mechanism. 

Extended Data Figure \ref{fig:ex_pair} (a) provides a simplified representation of the spin-pair model, consisting of two independent defects (Defect A and B), separated by $\ge$ 1 nm, forming a defect complex. This complex hosts two unpaired electrons that establish different internal charge states depending on their spatial occupancy.  When both electrons are localized on the same defect  (Defect A), they form a closed-shell spin singlet GS with a metastable spin triplet state (S = 1), which can be accessed through laser excitation and intersystem crossing (ISC) transitions (left panel in Extended Data Figure \ref{fig:ex_pair} (b)). This state corresponds to a strongly coupled  spin-pair charge state and explains the S = 1 transitions. 

Alternatively, laser excitation can induce charge hopping that transfers one electron from Defect A to Defect B, forming a weakly coupled defect pair. In this configuration, each defect hosts a single electron (S = 1/2), as illustrated in the right panel of Extended Data Figure \ref{fig:ex_pair} (b). This spin-dependent charge hopping yields a corresponding spin-dependent PL signal \cite{robertson2024universal}.

The actual GS, which is also the optically active state, is determined by the lowest-energy charge configuration and depends on the specific defect species and the local Fermi energy level. We considered two possible energy level configurations (Supplementary Section XII), with Extended Data Figure \ref{fig:ex_pair}(c) depicting the most likely scenario. In the most stable charge state, both electrons occupy the same defect site, yielding an S = 0 GS and a metastable S = 1 state. Laser excitation can then generate a metastable spin-pair charge state by promoting transitions from the S = 1 to the S = 1/2 manifold. The pronounced asymmetry in the Rabi oscillations, reflected by an increasing contrast baseline in both S = 1/2 and S = 1 transitions, strongly supports the metastable nature of these spin manifolds.

In the presence of a $^{13}$C nuclear spin, the defect electron spins can couple to the nuclear spin via hyperfine interactions,  which vary across different charge states.  In the weakly coupled spin-pair state, the nuclear spin is primarily coupled to a single electron spin at Defect A.  This is consistent with our experimental observations, where hyperfine coupling constants A$_{zz}$ of 130 MHz and 300 MHz were measured for Group II and Group III defects, respectively. 

Given their distinct hyperfine features, Group II and Group III defects are likely to have more well-defined and deterministic structures. In contrast, Group I defects, lacking hyperfine splitting, may include a range of chemical configurations: either similar to Group II and III defects (but involving $^{12}$C instead of $^{13}$C) or very different chemical structures. This structural variability may account for the broader range of stability observed in Group I defects, whereas Group II and III defects tend to exhibit more consistent and stable behavior under experimental conditions.

\vspace{0.1in}
\textbf{Simulation of spin photodynamics}

Based on the possible energy level models, we numerically simulate the spin photodynamics using the Lindblad master equation:
\begin{equation}
	\dot{\rho} = -i [H,\rho(t)]+\sum_{k}\Gamma_k\left[L_k\rho(t)L_k^\dagger-\frac{1}{2}\left\{L_k^\dagger L_k,\rho(t)\right\}\right], \label{eq:metastablepair}
\end{equation} 
where $\rho(t)$ is the time-dependent density matrix, $\Gamma_k$ represents transition rates, and $L_k$  are the associated Lindblad operators. This simulation allows us to predict PL signals in both CW ODMR and pulsed Rabi experiments (see details in Supplementary Information). 

In the CW ODMR simulation, we use the full Hamiltonian, which includes the optical manifold, spin-pair states, and spin-triplet states. For Model 1, where the GS is a spin singlet (S = 0), the Hamiltonian is written as:
\begin{equation}
	H = {H_{pair}} \oplus {H_{m,S1}} \oplus {H_{eg}} = \left( {\begin{array}{*{20}{c}}
			{H_{pair}^{8 \times 8}}&{}&{}\\
			{}&{H_{m,S1}^{6 \times 6}}&{}\\
			{}&{}&{H_{eg}^{4 \times 4}}
	\end{array}} \right)
\end{equation}
where $H_{eg}$, $H_{m,S1}$, and $H_{pair}$ describe the optical manifold, the spin S=1 MS and the spin-pair state, respectively (see detailed expressions in Supplementary Information). For each subspace, we consider a $^{13}$C nuclear spin coupled via hyperfine interaction. The full Hamiltonian is a direct sum of individual spin manifolds, meaning there is no coherent interaction between them. Instead, they are connected via incoherent transitions described by the Lindblad operators $L_k$.




\section*{ACKNOWLEDGMENTS}
T.L. acknowledges the support by the Gordon and Betty Moore Foundation, grant DOI 10.37807/gbmf12259, and the National Science Foundation grant PHY-2409607. The investigation of the spin pair model was supported by the U.S. Department of Energy, Office of Science, Office of Basic Energy Sciences through QuPIDC Energy Frontier Research Center under award Number DE-SC0025620. Y.P.  acknowledges the support by the National Science Foundation under grant no. DMR2143233. The {\it ab initio} calculations used resources of the Scientific Data and Computing Center, a component of the Computational Science Initiative, at Brookhaven National Laboratory under Contract No. DE-SC0012704, the Lux Supercomputer at UC Santa Cruz, funded by NSF MRI Grant No. AST 1828315, the National Energy Research Scientific Computing Center (NERSC), a U.S. Department of Energy Office of Science User Facility operated under Contract No. DE-AC02- 05CH11231, and the Extreme Science and Engineering Discovery Environment (XSEDE), which is supported by National Science Foundation Grant No. ACI-1548562.

\section*{AUTHOR CONTRIBUTIONS}
T.L. and X.G. conceived and designed the project. X.G., S.V., K.S. and Y.J. built the setup. K.L. and S.Z. performed the {\it ab initio} calculations. P.J. fabricated microwave waveguides. S.V. and S.D. performed the ion implantation and thermal annealing on the hBN samples. X.G. and S.V. performed measurements. X.G. and Z.G. performed the numerical simulation. X.G., S.V., T.L., K.L. and Y.P. analyzed the results.  T.L. and Y.P. supervised the project. All authors contributed to the writing of the manuscript.



\renewcommand{\figurename}{Extended Data Figure} 
\setcounter{figure}{0}

\begin{figure*}[!tph]
	\centering
	\includegraphics[width=0.7\textwidth]{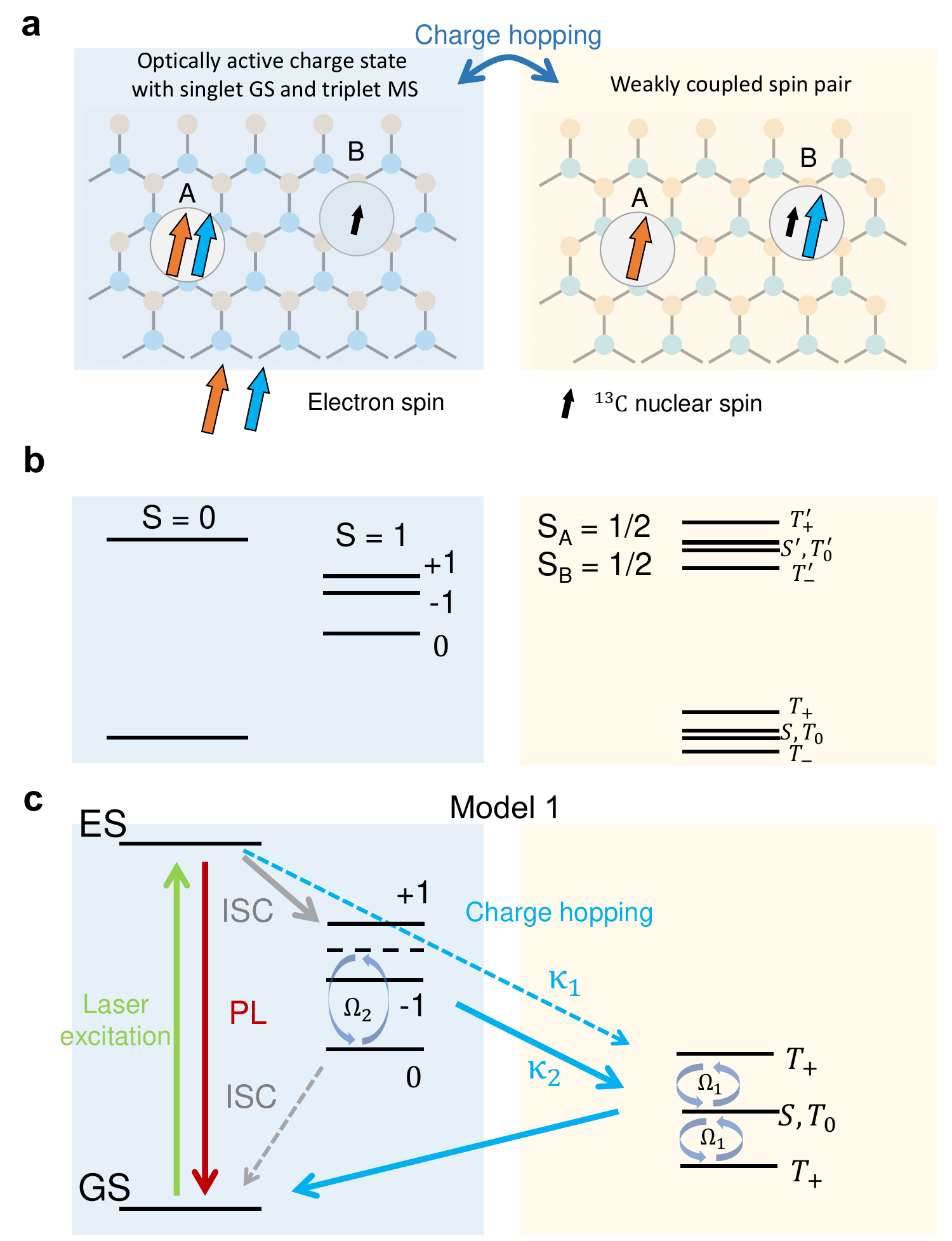}
	\caption{  \textbf{Spin pair model.} (a) Two possible electron configurations of a spin defect complex in different internal charge states.  (b) The energy levels of internal charge states according to the DFT prediction. When two electrons occupy the same defect site (left panel), it forms a singlet GS and ES as well as a triplet MS. When two electrons locate at different defect sites (right panel), each form a S = 1/2 state and they are weakly coupled via dipolar interaction, forming a spin S = 1/2 pair. (c) A proposed energy level diagram for explaining the coexistence of the S = 1/2 and S = 1 ODMR transitions. } \label{fig:ex_pair}
\end{figure*}

\end{document}